\documentclass[ reprint, amsmath,amssymb, prl,showpacs,superscriptaddress,]{revtex4-1}
\pdfoutput=1
\newcommand{\be}{\begin{eqnarray}}
\newcommand{\ee}{\end{eqnarray}}

\usepackage[pdftex]{graphicx}
\usepackage{epstopdf}

\newcommand{\bra}[1]{\langle{#1}|}
\newcommand{\ket}[1]{|{#1}\rangle}
\usepackage{graphicx}
\usepackage{dcolumn}
\usepackage{bm}

\begin{document}

\preprint{APS/123-QED}
\title{Single-Shot Quantum Non-Demolition Measurement of a Quantum Dot Electron Spin, using Cavity Exciton-Polaritons.}
\author{Shruti Puri}
\thanks{These authors contributed equally}
\affiliation{1. E. L. Ginzton Laboratory, Stanford University, Stanford, California 94305, USA}
\author{ Peter L. McMahon}
\thanks{These authors contributed equally}
\affiliation{1. E. L. Ginzton Laboratory, Stanford University, Stanford, California 94305, USA}
\author{Yoshihisa Yamamoto }
\affiliation{1. E. L. Ginzton Laboratory, Stanford University, Stanford, California 94305, USA}
\affiliation{2. National Institute of Informatics, 2-1-2 Hitotsubashi, Chiyoda-ku, Tokyo 101-8430, Japan}%
\date{\today}

\begin{abstract}
We propose a scheme to perform single-shot quantum non-demolition (QND) readout of the spin of an electron trapped in a semiconductor quantum dot (QD). Our proposal relies on the interaction of the QD electron spin with optically excited, quantum well (QW) microcavity exciton-polaritons. The spin-dependent Coulomb exchange interaction between the QD electron and cavity polaritons causes the phase and intensity response of left circularly polarized light to be different to that of the right circularly polarized light, in such a way that the QD electron's spin can be inferred from the response to a linearly polarized probe. We show that, by careful design of a sample with coupled QD and QW, it is possible to eliminate spin-flip Raman transitions. Thus, a QND measurement of the QD electron spin can be performed within a few 10's of nanoseconds with fidelity $\sim 99.9\%$. This improves upon current optical QD spin readout techniques across multiple metrics, including speed and scalability.
\end{abstract}

\maketitle

\begin{figure}
\includegraphics[width=5.5cm,height=3cm]{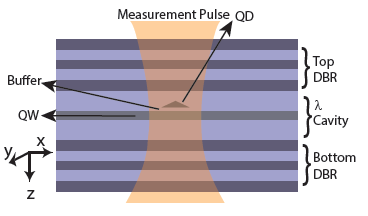}\\
\hspace{.25cm} (a) \\
\includegraphics[width=4cm,height=3.5cm]{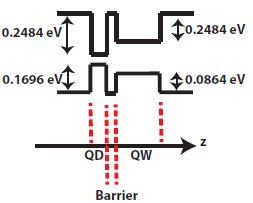}\includegraphics[width=4cm,height=3.5cm]{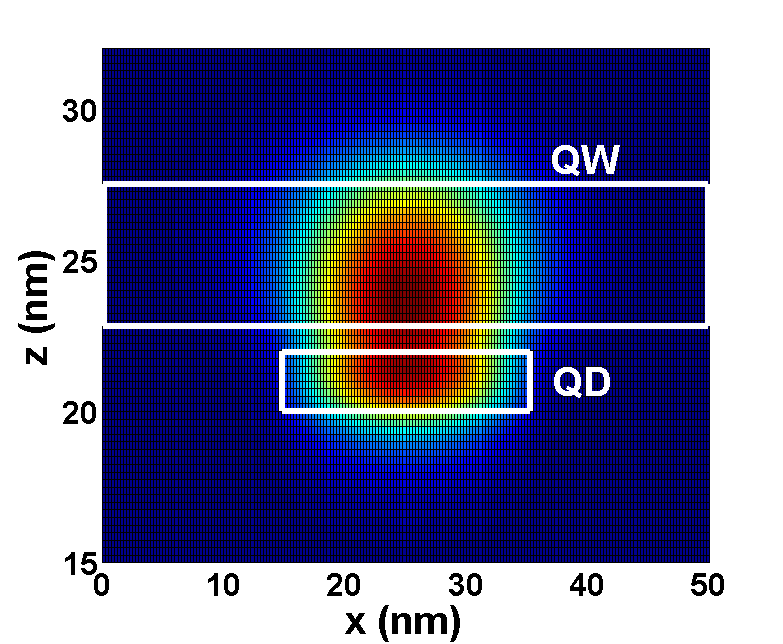}\\
(b)\hspace{3.5cm}(c)\\
\caption{(a) Illustration of the system consisting of a QD grown on a QW placed in a GaAs $\lambda$ microcavity. An electron is trapped in the QD. A probe laser incident over the QD, excites polaritons in the microcavity. (b) Band structure of the QD and QW system. The effective mass of electron (hole) in In$_{0.3}$Ga$_{0.7}$As is 0.0504$m_0$ (0.48$m_0$) and that in In$_{0.15}$Ga$_{0.85}$As is 0.0566$m_0$ (0.495$m_0$), where $m_0$ is the mass of a free electron. (c) Numerically evaluated, normalized wavefunction distribution ($|\psi(\bf{r})|^2$) of the QD electron along $x$ and $z$ axis at $y=25$ nm (the Schr\"{o}dinger equation is discretized in a cuboid region of size 50 nm $\times$ 50 nm$\times$ 50 nm. The white rectangles mark the QD and QW regions.   }
\end{figure} 

The ability to measure a single electron spin by projecting it onto the eigenstate corresponding to the measurement result, constitutes a QND measurement and is of great importance in measurement-based quantum computing schemes \cite{divin}. Since the electron spin is projected onto an eigenstate, the measurement can be repeated several times and should give the same result for subsequent measurements. Thus, classical noise can be reduced by time averaging. This method can be used for faithful initialization and measurement of qubits \cite{cody}.
\\
\indent
Any proposed QND spin measurement scheme should use a physical process that is unlikely to cause a spin-flip event for the duration of the measurement.  In an optical measurement scheme based on the Faraday rotation induced by a confined spin, the spin-flip Raman scattering must be supressed \cite{atature,berez}. One way to overcome this adverse effect is to use a QD molecule which has separate optical transitions for state preparation, manipulation and measurement \cite{vami}. However, even in this system the probe laser has a non-negligible probability of causing the spin to flip ($\approx 7\%$ in \cite{vami}). Furthermore, the measurement is quite slow, taking $\approx$ 2 ms to achieve a fidelity of 96$\%$. Finally, the use of QD molecules in a large scale quantum computing system has the disadvantage that it is difficult to deterministically grow arrays of spectrally homogeneous QD molecules.
\\
\indent
We propose a QND readout scheme for QD electron spins in Faraday geometry, using optically-excited QW exciton-polaritons. In Faraday geometry, a QD electron spin is quantized along the growth ($z$) axis, by an external magnetic field $B\hat{z}$. The system, illustrated in Fig.1(a,b), consists of a In$_x$Ga$_{1-x}$As QD grown on top of a In$_y$Ga$_{1-y}$As QW. Between them is a few monolayer thick GaAs barrier layer. The QD and QW are embedded in a GaAs $\lambda$ cavity formed by AlGaAs/AlAs distributed Bragg mirrors (DBRs). The QW exciton is resonant with the cavity photons at ${\bf{k}}_{||}={\bf{0}}$. In the strong coupling regime, bare QW excitons and cavity photons coherently exchange energy faster than the rate at which the photons are lost from the cavity. The resulting eigenmodes are upper polaritons (UPs) and lower polaritons (LPs), as depicted in Fig.2(a) \cite{polaritons}. The splitting between them depends on the strength of the coupling between QW excitons and cavity photons. 
A red-detuned ($\delta$), left(right) circularly polarized $\{\sigma^+(\sigma^-)\}$ laser pulse excites LPs in the region below the QD, as shown in Fig.1(a).  Because of the QW exciton selection rules, LPs with $J=+1(-1)$ and ${\bf{k}}_{||}={\bf{0}}$ are excited in the area ($A$) under the laser spot \cite{deng}. The excitonic component of the LP is composed of an electron with $s_{\textnormal{ze}}=-\frac{1}{2}(+\frac{1}{2})$ and a heavy-hole with $l_{\textnormal{zhh}}=+1(-1)$, $s_{\textnormal{zhh}}=+\frac{1}{2}(-\frac{1}{2})$, where $s$ and $l$ refer to spin and orbital angular momentum \cite{shel}. Excitons with $s_{\textnormal{ze}}=+\frac{1}{2}(-\frac{1}{2})$, $l_{\textnormal{zhh}}=+1(-1)$ and $s_{\textnormal{zhh}}=+\frac{1}{2}(-\frac{1}{2})$ are optically dark states. If the duration of the laser pulse is much longer than the inverse of optical detunings in the system, then the polaritons evolve adiabatically according to: 
\be 
\alpha_{1(-1)}(t)=\frac{\sqrt{\gamma_1}t_0f(t)_{1(-1)}}{i\delta+\frac{(\gamma_1+\gamma_2)}{2}}.\nonumber
\ee
Here, $\alpha_{1(-1)}$ are the coherent amplitudes of the LP with $J=1(-1)$, $|f(t)_{1(-1)}|^2$ are the input photon fluxes in $\sigma^+$($\sigma^-$) polarizations, $\gamma_1(\gamma_2)$ are the polariton decay rates from the top (bottom) DBR mirror, and $t_0$ is the photon Hopfield factor for LPs \cite{polaritons}. \\
\begin{figure}
\includegraphics[width=8cm,height=4cm]{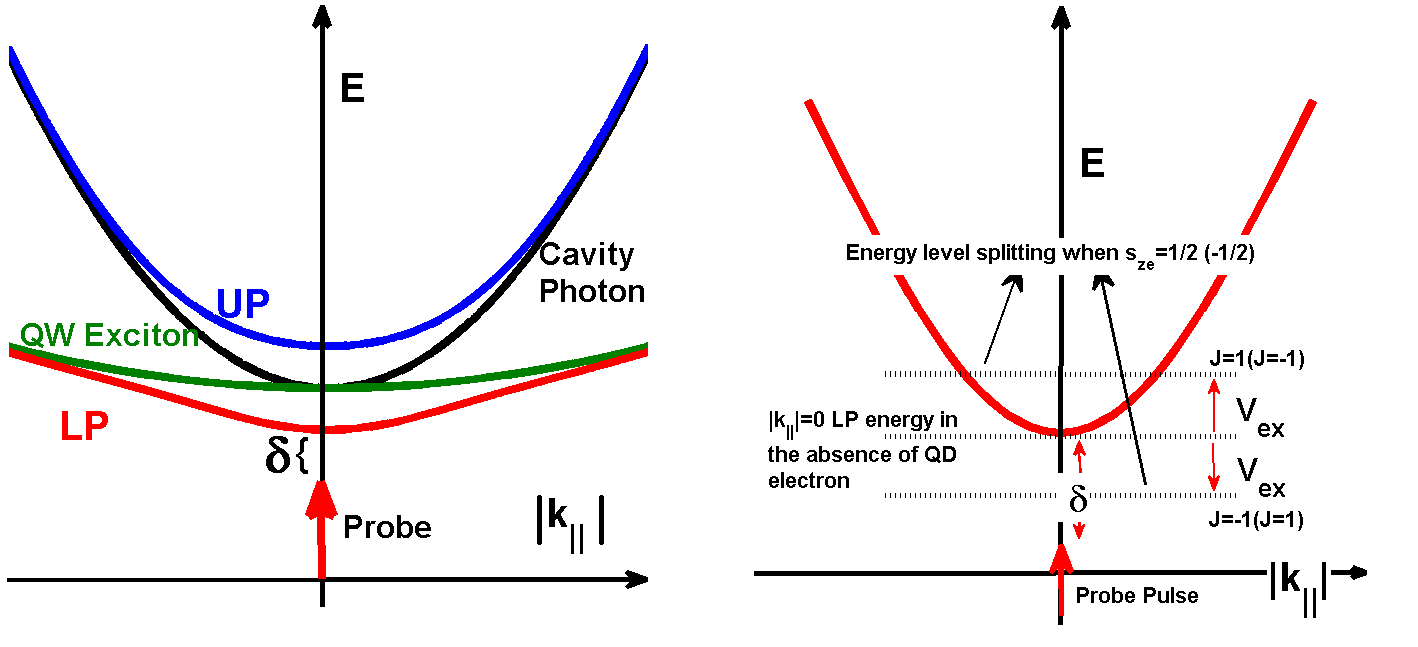}\\
(a)\hspace{3.8cm}(b)
\caption{(a) Representation of the exciton polariton energy dispersion. The green(black) line is the bare exciton(cavity photon) dispersion curves and blue(red) solid lines are that for the UPs(LPs).(b) Exaggerated depiction of the energy level splitting of the $J=1$ and $J=-1$ LP modes. If the QD electron spin $s_{\textnormal{ze}}=\frac{1}{2}(-\frac{1}{2})$, then the energy of the $J=-1$ polariton is red-detuned (blue-detuned) from the $J=1$ polariton by 2$V_{\textnormal{ex}}$.}
\end{figure}
\indent
A self-assembled In$_x$Ga$_{1-x}$As QD has a pyramidal shape with a typical height of $\sim1.5$ nm and base width of $\sim$ 20 nm. An In$_y$Ga$_{1-y}$As QW can be grown 6 nm thick. By carefully choosing the barrier layer thickness and In concentrations ($x$ and $y$) one can design the band structure such that the electron confined in the QD has a non-zero wavefunction in the QW. The finite overlap of the localized QD electron and microcavity polariton results in a spin-dependent Coulomb exchange interaction between them \cite{qui2,qui,puri}:
\be 
H_\textnormal{I}&=&-V_{\textnormal{ex}} {\hat{\sigma}_\textnormal{l}}\cdot {\hat{\sigma}_\textnormal{e}},\nonumber\\
|V_{\textnormal{ex}}|&=&|r_0|^2\int d{\bf{r}}_\textnormal{e}d{\bf{r}}_\textnormal{h}d{\bf{r}}_\textnormal{l}\frac{\psi({\bf{r}}_\textnormal{e},{\bf{r}}_\textnormal{h})\phi({\bf{r}}_\textnormal{l})e^2\psi({\bf{r}}_\textnormal{l},{\bf{r}}_\textnormal{h})\phi({\bf{r}}_\textnormal{e})}{4\pi\epsilon(|{\bf{r}}_\textnormal{e}-{\bf{r}}_\textnormal{l}|)}, \nonumber
\ee
where $\epsilon$ is the dielectric constant of the In$_y$Ga$_{1-y}$As QW, ${\bf{r}}_\textnormal{e},{\bf{r}}_\textnormal{h}$ are the position vectors of the electron and hole in the excitonic part of the polariton, ${\bf{r}}_\textnormal{l}$ is that of the localized electron, $\psi$ and $\phi$ represent the wavefunctions of the excitonic component of the polariton and of the localized electron, ${\hat{\sigma}_\textnormal{l}}$ (${\hat{\sigma}_\textnormal{e}}$) are the Pauli spin operators of the localized electron (electronic part of polariton). $r_0$ is the excitonic Hopfield coefficient of ${\bf{k}}_{||}={\bf{0}}$ LPs. Since the cavity photons and QW excitons are resonant at ${\bf{k}}_{||}={\bf{0}}$, $r_0=1/\sqrt{2}$. The unique area ($A=\pi R^2$) in the QW in which the polaritons are coherently excited depends on the cavity lifetime ($\tau$) \cite{yam, yam2, dbr}. For example, for the cavity photon lifetime $\tau=4$ ps, $R=3.6$ $\mu$m \cite{supl}. If $x=30\%$, $y=15\%$, the barrier layer is 1 nm thick and the pump laser excites the LPs in a spot of radius $R=3.6$ $\mu$m, then we estimate that $V_{\textnormal{ex}}\approx 0.2$ $\mu$eV \cite{supl}. We have designed a sample for which the band structure is such that the QD trion resonance (937 nm) is detuned from the QW exciton resonance (918 nm) by $\sim$ 27 meV. This ensures that the probe pulse, which has a frequency near that of the QW exciton resonance, is far detuned from the s-, p-, and higher-shell QD trion resonances. This results in a very low probability for the probe pulse to cause a QD spin-flip by Raman scattering. \\
\indent
The exchange interaction gives rise to not only spin-conserving but also spin-flip terms. The spin-conserving term induces a spin-dependent shift in the polariton resonance. If ${s}_{\textnormal{zl}}=+\frac{1}{2}$, then the resonance energy of a $J=-1(+1)$ LP will decrease (increase) by an amount $V_{\textnormal{ex}}$, making the $J=1$ and $J=-1$ polaritons non-degenerate as shown in Fig. 2(b). (This effect is reversed if ${s}_{\textnormal{zl}}=-\frac{1}{2}$.) We will exploit these spin-dependent polariton resonance shifts to measure the spin of the QD electron. If the localized electron undergoes a spin-flip, the LP will be scattered as a dark exciton. The dark exciton is blue-detuned by $\Delta_{\textnormal{dark}}\sim$ 1 meV from the LPs at ${\bf{k}}_\|={\bf {0}}$ and thus this scattering is made possible only by phonon absorption. We  show in \cite{supl} that the spin-flip probability in our proposed scheme is negligible. The total Hamiltonian of the system can be written as:
\be 
H&=&\delta p^\dag_1 p_1 + \delta p^\dag_{-1} p_{-1}-V_{\textnormal{ex}}\hat{\sigma}_{\textnormal{ze}}p^\dag_1 p_1+V_{\textnormal{ex}}\hat{\sigma}_{\textnormal{ze}}p^\dag_{-1} p_{-1}\nonumber\\
&+&\sqrt{\gamma_1}f_{\textnormal{1in}}(p^\dag_1+p_1)+\sqrt{\gamma_1}f_{\textnormal{-1in}}(p^\dag_{-1}+p_{-1}),\nonumber
\ee
where $\delta$ is the detuning of the $J=1$ and $J=-1$ polaritons from the probe pulse in the absence of the exchange interaction (shown in Fig. 2(a,b)), $\hat{\sigma}_{\textnormal{ze}}=\left(\left |\frac{1}{2}\right \rangle \left\langle \frac{1}{2}\right |-\left |-\frac{1}{2} \right\rangle\left\langle -\frac{1}{2}\right|\right)$ is the Pauli spin operator ($\left |\pm \frac{1}{2}\right \rangle$ is the spin state of the localized electron), $p^\dag_{1}(p^\dag_{-1})$ are the creation operators of $J=1(-1)$ polaritons, and $|f_{\textnormal{1in}}|^2(|f_{-\textnormal{1in}}|^2)$ is the polariton flux i.e., the number of polaritons that are pumped into the QW per unit time. 
\begin{figure}
\includegraphics[width=7.25cm,height=3.75cm]{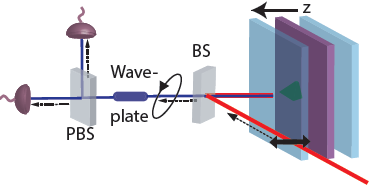}
\caption{Illustration of the measurement setup.  }
\end{figure} 
\\
\indent
The setup for the measurement scheme is illustrated in Fig. 3. A horizontally (H) polarized probe laser pulse is incident on the coupled QD-QW microcavity system (through a 90:10 beam splitter),  coherently exciting polaritons with $J=1$ and $J=-1$. These polaritons interact with the localized spin and decay from the cavity at rate $\gamma=\gamma_1+\gamma_2$, emitting $\sigma^+$- and $\sigma^-$-polarized photons respectively. Because the $J=1,-1$ polaritons evolve with different frequencies depending on the QD electron spin, the light coupled out from the cavity carries information about the QD spin. As a result, the reflected light is elliptically polarized with its axis tilted by an angle $\propto \pm V_{\textnormal{ex}}$ (depending on whether $s_{\textnormal{ze}}=\pm \frac{1}{2}$). Even in the absence of the QD electron, strain-induced asymmetry during the growth process can lift the degeneracy between H- and V-polarized polaritons \cite{kal,geo}. Considering this energy splitting ($=2V_\textnormal{s}$), the photon flux reflected from the cavity is:
\be 
\frac{f_\textnormal{H}}{f_0}&=&-1+\frac{\gamma_1\left(i\delta_2+\frac{\gamma}{2}\right)}{V^2_{\textnormal{ex}}+\left(i\delta_1+\frac{\gamma}{2}\right)\left(i\delta_2+\frac{\gamma}{2}\right)}\nonumber\\
\frac{f_\textnormal{V}}{f_0}&=&\frac{\mp \gamma_1 V_{\textnormal{ex}}}{V^2_{\textnormal{ex}}+\left(i\delta_1+\frac{\gamma}{2}\right)\left(i\delta_2+\frac{\gamma}{2}\right)}.\nonumber
\ee
Here, $|f_0|^2$ is the H-polarized input photon flux, $|f_\textnormal{H}|^2(|f_\textnormal{V}|^2)$ is the reflected photon flux with H (V) polarization, with the - or + indicating if $s_\textnormal{zl}=+ \frac{1}{2}$ or $ -\frac{1}{2}$. $\delta_1(\delta_2)$ is the detuning of the laser from the H (V)-polarized LPs in the absence of a QD electron, so that $\delta_1-\delta_2=2V_{\textnormal{s}}$. The reflected light from the cavity passes through the 90:10 BS and arrives at the waveplate. The $\frac{\lambda}{2}(\frac{\lambda}{4})$ waveplate, with its axis oriented at $\frac{\pi}{4}(0)$ rad with respect to the H-V axis rotates the polarization of the field. The polarizing beam splitter (PBS), placed at the output of the $\frac{\lambda}{2}(\frac{\lambda}{4})$ waveplate, oriented along (45$^\circ$ to) the axis of the waveplates, isolates the two orthogonal polarizations incident on it, which are then measured by detectors D$_1$ and D$_2$. The difference in the photon counts of D$_1$ and D$_2$, when using a $\frac{\lambda}{2}$ waveplate is:
\be 
I_{\textnormal{D1}}-I_{\textnormal{D2}}&=&\left |\frac{f_\textnormal{H}+f_\textnormal{V}}{\sqrt{2}}\right |^2-\left |\frac{f_\textnormal{H}-f_\textnormal{V}}{\sqrt{2}}\right |^2\nonumber\\
&=&2|f_{+}||f_{-}|\sin(\theta_{+}-\theta_{-}), 
\label{disp}
\ee
which is the phase response. When using a $\frac{\lambda}{4}$ waveplate the difference in detector counts is:
\be 
I_{\textnormal{D1}}-I_{\textnormal{D2}}&=&\left |\frac{f_\textnormal{H}+if_\textnormal{V}}{\sqrt{2}}\right |^2-\left |\frac{f_\textnormal{H}-if_\textnormal{V}}{\sqrt{2}}\right |^2\nonumber\\
&=&{|f_+|^2-|f_{-}|^2},
\label{abs}
\ee
which is the intensity response. In the above equations, $f_+(f_{-})$ and $\theta_+(\theta_{-})$ are the amplitudes and phase shifts of the reflected field with $\sigma^+(\sigma^-)$ polarization \cite{supl}. In  both cases $I_{\textnormal{D1}}-I_{\textnormal{D2}}\propto \pm V_\textnormal{ex}$ (for small $V_{\textnormal{ex}}$) if $s_\textnormal{zl}=\pm \frac{1}{2}$ and hence can be used to distinguish the spin state of the localized electron spin. \\
\begin{figure}
\includegraphics[width=4.5cm,height=4.25cm]{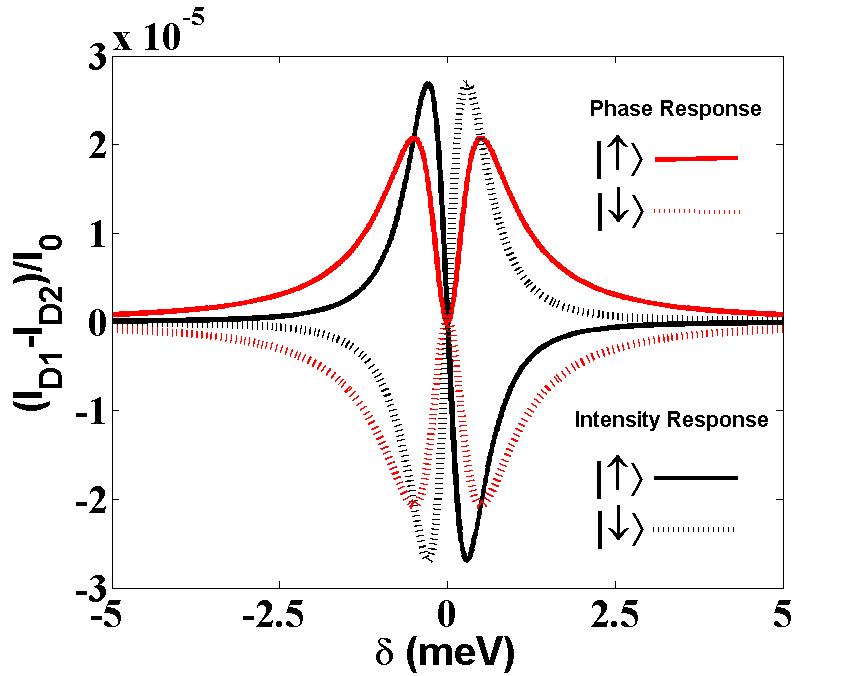}\includegraphics[width=4.5cm,height=4.25cm]{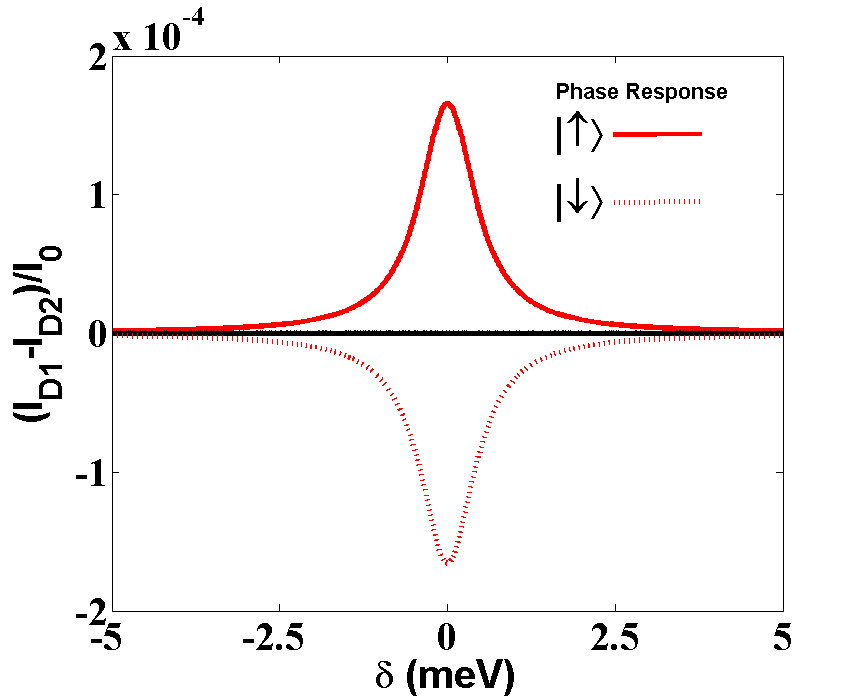}
\\
\hspace{1cm} (a)\hspace{4cm} (b)\\
\caption{Phase (red) and intensity (black) response when $V_\textnormal{s}=0$, (a) $\gamma_1=\gamma_2=0.5$ meV, (b) $\gamma_1=1$ meV, $\gamma_2=0$. The solid (dashed) lines represent the response with the QD electron spin is $s_{\textnormal{zl}}=\frac{1}{2}(-\frac{1}{2})$. }
\end{figure} 
\indent 
Figure 4 shows the phase and intensity responses in the reflection spectrum of the cavity. If $s_{\textnormal{zl}}=+\frac{1}{2}$,  $V_\textnormal{s}=0$, then $\delta_1=\delta_2=\delta$. For $\delta<0$ ($\delta>0$), the probe pulse is closer to the $J=-1$ ($J=1$) LP resonance (Fig. 2). As a result, in a two-sided cavity, when $\delta<0$ ($\delta>0$)  $\sigma^+$-polarized light (which excites $J=1$ LPs) will be reflected more (less) than the $\sigma^-$ light (which excites $J=-1$ LPs). If $|\delta|\gg \gamma$, then neither of the polarization components of the probe pulse are able to enter the cavity. Consequently, there is no information about the spin state of the QD electron in the reflected light and from Eqn. \ref{abs}, $I_{\textnormal{D1}}-I_{\textnormal{D2}}=0$. At $\delta\approx\pm \gamma/(2\sqrt{3})$ the intensity response becomes maximal. These  results can be seen in the intensity response shown in Fig. 4(a). On the other hand, in a single-sided cavity, for all values of detuning $\delta$, both the $\sigma^+$ and $\sigma^-$ light are completely reflected from the cavity. Hence, the intensity response vanishes (Fig. 4(b)). The phase response from a two sided cavity can be understood as follows: at $\delta=V_{\textnormal{ex}}(-V_{\textnormal{ex}})$, the probe is resonant with the $J=1(-1)$ polariton mode. Hence, $f_+(f_-)=0$ and from Eqn. \ref{disp} for the dispersive response, $I_{\textnormal{D1}}-I_{\textnormal{D2}}=0$. In a two sided cavity, $\tan(\theta_+)=\frac{\gamma}{2(\delta+V_{\textnormal{ex}})}$ and $\tan(\theta_-)=\frac{\gamma}{2(\delta-V_{\textnormal{ex}})}$. As shown in Fig. 4(a), the maximum in the phase response appears at $\delta\approx \pm \gamma/2$. In a single sided cavity $\tan(\theta_+)=4(\delta+V_{\textnormal{ex}})/\gamma$ and $\tan(\theta_-)=4(\delta-V_{\textnormal{ex}})/\gamma$. Its phase response is shown in Fig. 4(b). Unlike the two-sided cavity, the phase response of a single-sided cavity is maximal at $\delta=0$. If $s_{\textnormal{zl}}=-\frac{1}{2}$ then the $J=1$ polariton mode has lower energy than the $J=-1$ mode and the response curves are just reversed (dotted lines in Fig. 4(a,b)). Thus a measurement $I_{\textnormal{D1}}-I_{\textnormal{D2}}$ will reveal the spin state of the electron.
 In a real experimental system $V_{\textnormal{s}}\neq 0$ and in \cite{supl} we have plotted and explained the response curves for a typical H-V nondegeneracy of $V_\textnormal{s}=0.15$ meV \cite{kal,geo}.  \\
\indent
As explained previously, one source of  erroneous operation in this measurement scheme is the phonon-assisted, spin-flip scattering and its probability is $P_\textnormal{e}^{\textnormal{dark}}\sim \Gamma^\textnormal{dark} \tau_{\textnormal{meas}}$ where $\Gamma^\textnormal{dark}=63300$ s$^{-1}$ (418 s$^{-1}$) at $T=1.5$ K for a single-sided (two-sided) cavity \cite{supl}. In addition, since we optically excite $N$ exciton-polaritons and the QD electron can radiatively recombine with any of the $N$ hole states in the QW. As we explain in \cite{supl}, the oscillator strength of such a transition is very small, leading to a long radiative lifetime $\tau_0$ ($\sim$100 ms). The probability of error during the measurement time $\tau_{\textnormal{meas}}$ is $P^{\textnormal{rad}}_e$= $1-e^{-N\tau_{\textnormal{meas}}/\tau_0}$. Finally, in a photon counting measurement, there are errors due to quantum fluctuations (shot noise). The number of polaritons at steady-state is limited to $N\sim 2000$, corresponding to a low density of $\sim 5\times 10^{9}$ cm$^{-2}$, so that, any polariton-polariton interaction can be ignored \cite{supl}. The measurement time is set by the amount of time needed to probe the system with sufficiently many photons that the maximal signal (in Fig. 3) can be reliably used to discriminate $s_{\textnormal{ze}}=+\frac{1}{2}$ and $s_{\textnormal{ze}}=-\frac{1}{2}$. Table 1 shows the measurement times required to make measurements that have a discrimination error of $P^{\textnormal{sn}}_{\textnormal{e}}=4\times 10^{-4}$ due to shot noise. One can measure the spin state of the electron spin qubit within 28 ns (for $V_\textnormal{s}=0$) or 17 ns (for $V_\textnormal{s}=0.15$ meV), with overall fidelities of $\sim 99.95\%$. A single-shot measurement taking only $\tau_{\textnormal{meas}}\sim 10$ ns would represent a $10^5$-fold improvement in speed over the current best single-shot readout \cite{vami}. \\
\indent
In conclusion, we have predicted that it is possible to perform a single-shot QND readout of the spin state of a QD electron by measuring the phase or the intensity response of a linearly polarized probe laser reflected from a cavity in which a QD  is embedded close to the QW.\\
\indent
This work has been supported by the Japan Society for the Promotion of Science (JSPS) through its ``Funding Program for World-Leading Innovative R$\&$D on Science and Technology (FIRST Program)" and NICT. We thank Prof. C. Piermarocchi, Dr. S. H\"{o}fling, Dr. C. Schneider, Prof. D. Miller, Prof. W. Harrison,  Dr. M. Fraser and Dr. T. Byrnes for insightful discussions.
\vspace{-.825cm}
\begin{widetext}
\begin{center}
\vspace{-.4cm}
\begin{table}[h]
\caption{Required measurement time $\tau_{\textnormal{meas}}$ assuming $\gamma=\gamma_1+\gamma_2=1$ meV, $V_{\textnormal{ex}}=0.2$ $\mu$eV,  $P^{\textnormal{sn}}_\textnormal{e}=0.04\%$, $P^{\textnormal{dark}}_\textnormal{e}\sim N\Gamma \tau_\textnormal{meas}$, $P^{\textnormal{rad}}_{\textnormal{e}}\sim N\tau_{\textnormal{meas}}/\tau_0$ ($\tau_0\sim 100$ ms) and $P^{\textnormal{total}}_{\textnormal{e}}=P^{\textnormal{sn}}_\textnormal{e}+P^{\textnormal{dark}}_\textnormal{e}+P^{\textnormal{rad}}_{\textnormal{e}}.$ All $P_\textnormal{e}$'s are listed in $\%$ and $\tau_\textnormal{meas}$ is listed in ns.}
\begin{tabular}{||c|c|c|c|c||}
\hline
\hline
Response of two-sided cavity & $\tau_{\textnormal{meas}}$  & $P^{\textnormal{rad}}_{\textnormal{e}}$&$P^{\textnormal{dark}}_{\textnormal{e}}$ & $P^{\textnormal{total}}_{\textnormal{e}}$ \\ \hline
Phase ($V_{\textnormal{s}}=0$) & 64 & 0.028 &0.0028 & 0.045\\ \hline
Intensity ($V_{\textnormal{s}}=0$) & 28 & 0.01 &0.001& 0.05\\ \hline
Phase ($V_{\textnormal{s}}=0.15$ meV) & 72 & 0.04 & 0.003& 0.08\\ \hline
Intensity ($V_{\textnormal{s}}=0.15$ meV) & 17 & 0.009 &0.0007& 0.05 \\ \hline
\end{tabular}
\begin{tabular}{||c|c|c|c|c||}
\hline
\hline
Response of single-sided cavity & $\tau_{\textnormal{meas}}$ & $P^{\textnormal{rad}}_{\textnormal{e}}$ &$P^{\textnormal{dark}}_{\textnormal{e}}$ & $P^{\textnormal{total}}_{\textnormal{e}}$ \\ \hline
Phase ($V_{\textnormal{s}}=0$) & 8 & 0.005&0.05 & 0.095\\ \hline
Intensity ($V_{\textnormal{s}}=0$) & - & - & -&-\\ \hline
Phase ($V_{\textnormal{s}}=0.15$ meV) & 12 & 0.006&0.07 & 0.1\\ \hline
Intensity ($V_{\textnormal{s}}=0.15$ meV) & 28 & 0.01&0.17 & 0.2 \\ \hline
\end{tabular}
\end{table}
\vspace{-.8cm}
\end{center}
\end{widetext}

\newpage

\preprint{APS/123-QED}
\title{Single-Shot Quantum Non-Demolition Measurement of Quantum Dot Electron Spins, using Cavity Exciton-Polaritons.}
\author{Shruti Puri}
\thanks{These authors contributed equally}
\affiliation{1. E. L. Ginzton Laboratory, Stanford University, Stanford, California 94305, USA}
\author{ Peter McMahon}
\thanks{These authors contributed equally}
\affiliation{1. E. L. Ginzton Laboratory, Stanford University, Stanford, California 94305, USA}
\author{Yoshihisa Yamamoto }
\affiliation{1. E. L. Ginzton Laboratory, Stanford University, Stanford, California 94305, USA}
\affiliation{2. National Institute of Informatics, 2-1-2 Hitotsubashi, Chiyoda-ku, Tokyo 101-8430, Japan}%
\date{\today}

\maketitle

\section{Wavefunction of electron qubit.}
When an electron is trapped in a QD adjacent to a QW, its wavefunction is spread in both the QD and the QW. We determine this wavefunction by numerically solving the time independent Schr\"{o}dinger equation in discretized space. The space is divided into a grid of $N\times N\times N$ elements, each with volume $\Delta \times \Delta \times \Delta$. If each element of the grid is denoted by $j$, then the discretized Hamiltonian is \cite{p1}:
\be 
H_{j,j}=V(r_j)+\frac{3}{\Delta^2}, \quad H_{\delta [j]j}=-\frac{1}{2}\frac{1}{\Delta^2}\nonumber
\ee
where, $V(r_j)$ is the potential at the element $j$ and $\delta [j]$ denotes an element nearing $j$, of which there are 6 in 3D. $ H_{\delta [j]j}$ represents the kinetic energy hops among nearest elements. Next, the Lanczos algorithm is applied \cite{p1}, following which, an initial vector $\ket{\phi_1}$ is chosen and an orthonormal basis $Q=\{\ket{\phi_1}, \ket{\phi_2},...,\ket{\phi_M}\}$ is generated. The Hamiltonian, written in the basis of $\ket{\phi_i}$'s is tridiagonal and can be easily diagonalized.  \\
\indent
The QD, composed of In$_{0.3}$Ga$_{0.7}$As, is approximated as a box of height 1.5 nm and base of 20 nm $\times$ 20 nm. Its base is parallel to the plane of a 6 nm thick QW, composed of In$_{0.15}$Ga$_{0.85}$As. The GaAs buffer between the QD and QW is 1 nm thick (Fig. 1(a)). The band structure of this system is shown in Fig. 1(b) \cite{b1}. The wavefunction of the electron, localized in the QD-QW region, found numerically, is shown in Fig. 1(c). We used $N=201$, $\Delta=0.25$ nm and convergence was obtained within $M=150$ iterations. From the numerical result, the wavefunction of the electron in the QW region can be approximated as:
\be 
\phi(\vec{r})=N_\textnormal{m} e^{-|z-z_0|^2/b^2}e^{-\rho^2/a^2}, \quad N_\textnormal{m}=0.0216 \textnormal{ (nm)}^{-3/2}\nonumber\\
\label{a}
\ee
where, $z$ is the position along the $z$ axis, $\vec{\rho}$ is the ($x$,$y$) position vector, $z_0=2$ nm, $b=4.7$ nm and $a=12$ nm. In the above equation, $z$ and $\vec{\rho}$ are measured from the center of the QW. The ground state energy, $E_0$ and first excited state energy, $E_1$ of the electron are 0.1612 eV and 0.1898 eV respectively. We propose to run the experiments at temperature $\sim 4$ K. As a result, $E_1-E_0=28.6$ meV $\gg k_\textnormal{B}T$, implying that the electron is indeed localized and can be employed as a qubit. A similar analysis is done for a hole, and the emission wavelength of a localized trion is obtained as 937 nm (assuming a binding energy of 20 meV \cite{beqd}). 
\begin{figure}
\includegraphics[width=4.5cm,height=4.5cm]{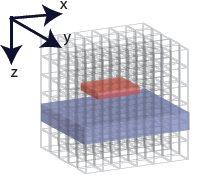}\includegraphics[width=4.5cm,height=4.5cm]{qndsuppl1.png}\\
\hspace{.25cm} (a) \hspace{4cm} (b)\\
\vspace{.5cm}
\includegraphics[width=6cm,height=5.5cm]{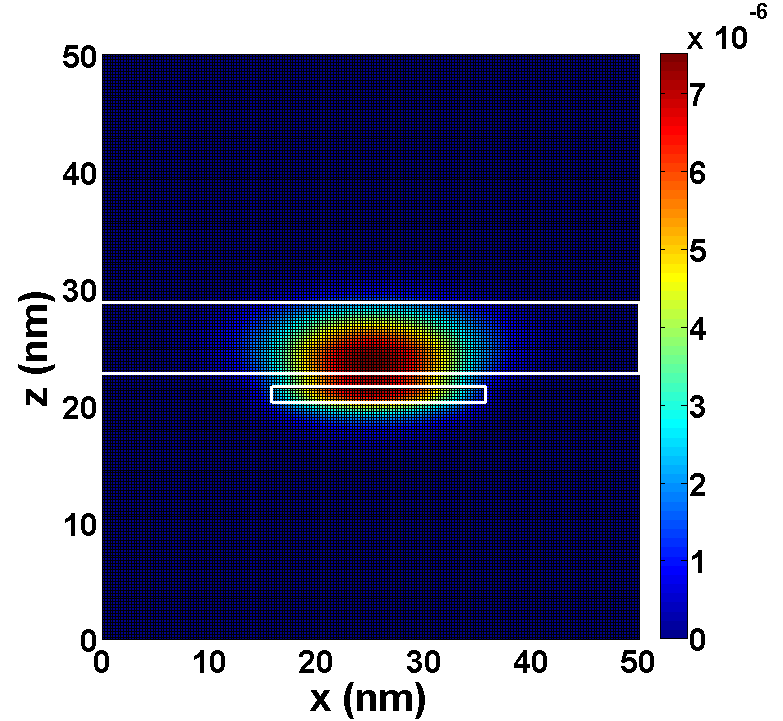}\\
(c)
\caption{(a) Illustration of the QD and QW structures in bulk GaAs in a region of dimensions 50 nm$\times$50 nm$\times$ 50 nm. (b) Band structure of the QD and QW system. The effective mass of electron (hole) in In$_{0.3}$Ga$_{0.7}$As is 0.0504$m_0$ (0.48$m_0$) and that in In$_{0.15}$Ga$_{0.85}$As is 0.0566$m_0$ (0.495$m_0$), where $m_0$ is the mass of a free electron. (c) Normalized wavefunction distribution of the electron along $x$ and $z$ axis at $y=25$ nm. The white rectangles mark the QD and QW regions.}
\end{figure}
\vspace{-.5cm}
\section{Wavefunction of QW exciton}
In QWs having thickness $L\ll a_\textnormal{B}$, where $a_\textnormal{B}$ is the Bohr radius of the exciton, the wavefunction of the exciton can be written as \cite{han}:
\be 
\psi(\vec{r}_i,\vec{}r_j)=\frac{e^{i{\bf{k}}_\|\cdot \vec{R}}}{\sqrt{A}}\sqrt{\frac{2}{\pi}}\frac{2}{a_\textnormal{B}}e^{-2|\vec{\rho}_i-\vec{\rho}_j|/a_\textnormal{B}}g(z_i)h(z_j)\nonumber
\ee
where $A$ is the area in which the excitons are excited, $\vec{R}$ (${\bf{k}}_\|$) is the center of mass coordinate (momentum) in the plane of the QW, $\vec{\rho}_i(\vec{\rho}_j)$ is the position of the electron(hole) in the plane of the QW and $g(z_i)$ ($h(z_j)$) is the electron (hole) wavefunction along the $z$ axis which can be determined by the numerical procedure outlined in the previous section. We will be exciting only $\bf{k}_\|={\bf{0}}$ excitons in the system and so:
\be 
\psi(\vec{r}_i,\vec{r}_j)&=&\frac{1}{\sqrt{A}}\sqrt{\frac{2}{\pi}}\frac{2}{a_\textnormal{B}}e^{-2|\vec{\rho}_i-\vec{\rho}_j|/a_\textnormal{B}}g(z_i)h(z_j)\nonumber\\
g(z_i)&=&M_1e^{-z_i^2/c_1^2},\quad M_1=0.4054 \textnormal{ (nm)}^{-1/2}\nonumber\\
h(z_j)&=&M_2e^{-z_i^2/c_2^2},\quad M_2=0.5 \textnormal{ (nm)}^{-1/2}
\label{b}
\ee 
where, $a_\textnormal{B}\sim 10$ nm, $c_1=6$ nm, $c_2=4$ nm. The QW exciton emission, assuming a binding energy of 5 meV \cite{beqw} is at 918 nm. This implies that the QD trion emission is red detuned by $\sim$ 27 meV from the QW exciton line. 
\vspace{-.45cm}
\section{Exchange Energy}
As explained in the paper, $V_\textnormal{ex}$, the exchange coupling energy between the localized electron spin and the electronic part of the polariton is given as:
\be 
V_\textnormal{ex}&=&\frac{1}{2}\int d{\bf{r}}_\textnormal{e}d{\bf{r}}_\textnormal{h}d{\bf{r}}_\textnormal{l}\frac{\psi({\bf{r}}_\textnormal{e},{\bf{r}}_\textnormal{h})\phi({\bf{r}}_\textnormal{l})e^2\psi({\bf{r}}_\textnormal{l},{\bf{r}}_\textnormal{h})\phi({\bf{r}}_\textnormal{e})}{4\pi\epsilon(|{\bf{r}}_\textnormal{e}-{\bf{r}}_\textnormal{l}|)}.\nonumber\\
\label{v}
\ee
The factor of $1/2$ is needed to take into account that only half of the polariton is excitonic, the other half being photonic (i.e., $|r_0|^2=\frac{1}{2}$). Substituting Eqns. \ref{a} and \ref{b} in Eqn. \ref{v}, followed by the transformation $\vec{\rho}_\textnormal{e}-\vec{\rho}_\textnormal{h}=\vec{t}$ and $\vec{\rho}_\textnormal{l}-\vec{\rho}_\textnormal{h}=\vec{s}$, gives
\be 
V_\textnormal{ex}=\frac{N_\textnormal{m}^2I_1I_2}{A\pi^2 \epsilon a_\textnormal{B}^2}\int \frac{e^{-|\rho_e+\vec{s}-\vec{t}|^2/a^2-\frac{2t}{a_\textnormal{B}}-\rho_e^2/a^2-\frac{2s}{a_\textnormal{B}}} }{|\vec{t}-\vec{s}|}d\vec{\rho}_\textnormal{l}d\vec{\rho}_\textnormal{e} d\vec{\rho}_\textnormal{h},\nonumber
\ee
where $I_i=M_i \int e^{-z^2/c_i^2}e^{-|z-z_0|^2/b^2} dz$. Next, let $\vec{\rho}_e+(\vec{s}-\vec{t})/2=\vec{u}$, $\vec{t}-\vec{s}=\vec{y}$, $\vec{t}+\vec{s}=\vec{x}$ and approximate $e^{x/\gamma}\approx e^{-x^2/2\gamma^2}$. Hence:
\be
V_\textnormal{ex}= \frac{N_\textnormal{m}^2I_1I_2a^2\pi\sqrt{\pi}\lambda}{8A\epsilon}, \quad 1/\lambda^2=1/2a^2+1/a_\textnormal{B}^2.
\ee

\section{Phase and Intensity Response when $V_{\textnormal{s}}= 0$}
Recall Eqns. 1 and 2 from the paper:
\be 
\frac{f_\textnormal{H}}{f_0}&=&-1+\frac{\gamma_1\left(i\delta_2+\frac{\gamma}{2}\right)}{V^2_{\textnormal{ex}}+\left(i\delta_1+\frac{\gamma}{2}\right)\left(i\delta_2+\frac{\gamma}{2}\right)}\nonumber\\
\frac{f_\textnormal{V}}{f_0}&=&\frac{-\gamma_1 V_{\textnormal{ex}}}{V^2_{\textnormal{ex}}+\left(i\delta_1+\frac{\gamma}{2}\right)\left(i\delta_2+\frac{\gamma}{2}\right)}.\nonumber
\ee
Thus the reflected field with $\sigma^+(\sigma^-)$ polarization, when $V_\textnormal{s}=0$, can be written as:
\be 
\frac{f_+}{f_0}&=&\frac{1}{\sqrt{2}}\left (-1+\frac{\gamma_1}{i(\delta-V_\textnormal{ex})+\frac{\gamma}{2}}\right ),\nonumber\\
\frac{f_-}{f_0}&=&\frac{1}{\sqrt{2}}\left (-1+\frac{\gamma_1}{i(\delta-V_\textnormal{ex})+\frac{\gamma}{2}}\right ).\nonumber
\ee
Thus the amplitudes and phase shifts of the reflected field from a single-sided cavity (i.e., $\gamma=\gamma_1$ and $\gamma_2=0$) are:
\be 
\frac{f_+}{f_0}&=&\frac{f_-}{f_0}=\frac{1}{\sqrt{2}},\nonumber\\
\tan(\theta_+)&=&\frac{\gamma (\delta+V_\textnormal{ex})}{\frac{\gamma^2}{4}-(\delta+V_\textnormal{ex})^2}, \quad \textnormal{and} \nonumber\\
\tan(\theta_-)&=&\frac{\gamma (\delta-V_\textnormal{ex})}{\frac{\gamma^2}{4}-(\delta-V_\textnormal{ex})^2}.\nonumber
\ee
The amplitudes and phase shifts of the reflected field from a symmetric two-sided cavity (i.e., $\gamma_1=\gamma_2=\gamma/2$) are:
\be 
\frac{f_+}{f_0}&=&\left |\frac{\delta+V_\textnormal{ex}}{\sqrt{2\left ((\delta+V_\textnormal{ex})^2+\frac{\gamma^2}{4}\right )}}\right |,\nonumber\\
\frac{f_-}{f_0}&=&\left |\frac{\delta-V_\textnormal{ex}}{\sqrt{2\left ((\delta-V_\textnormal{ex})^2+\frac{\gamma^2}{4}\right )}}\right |,\nonumber\\
\tan(\theta_+)&=&\frac{\gamma}{2(\delta+V_\textnormal{ex})}, \quad \textnormal{and} \nonumber\\
 \tan(\theta_-)&=&\frac{\gamma}{2(\delta-V_\textnormal{ex})}.\nonumber
\ee

\section{Phase and Intensity Response when $V_{\textnormal{s}}\neq 0$}
\begin{figure}
\hspace{-1.5cm}
\includegraphics[width=10cm,height=8cm]{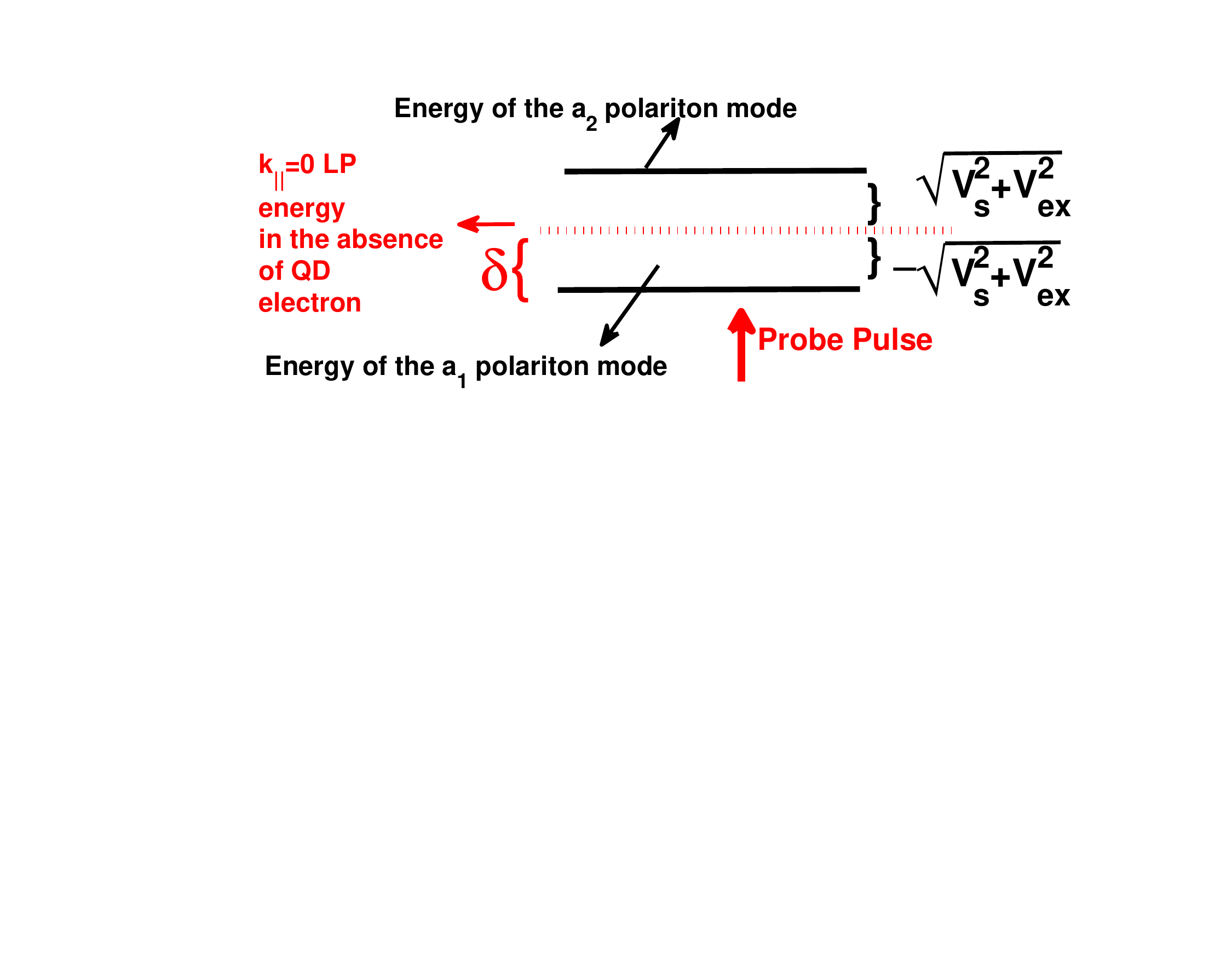}
\vspace{-5cm}
\caption{Energy level splitting of the new eigenstates, $a_1, a_2$ of the system when $V_\textnormal{s}\neq 0$}
\end{figure}
As was mentioned in the paper, even in the absence of a QD electron, the degeneracy between the H- and V-polarized  LPs is lifted due to growth-induced strain. Although the response curves can be easily plotted from equations in the main paper, in this section we give an intuitive understanding of the major differences between the case when $V_\textnormal{s}=0$ and $V_\textnormal{s}\neq 0$. In the latter case, the bare $J=-1$, $J=1$ polaritons are no longer the eigenstates of the Hamiltonian. The eigenmodes are the dressed states, separated in energy by $\sqrt{V_\textnormal{ex}^2+V_\textnormal{s}^2}\approx V_\textnormal{s}$ and formed by the linear combination (Fig. 2):
\be 
a_1&=& \frac{1}{\sqrt{2}}\left (1-\kappa\right )p_1+\frac{1}{\sqrt{2}}\left(1+\kappa\right )p_{-1},\nonumber\\
a_2&=& \frac{-i}{\sqrt{2}}\left (1+\kappa\right )p_1+\frac{i}{\sqrt{2}}\left (1-\kappa\right )p_{-1},
\label{mix}
\ee
where $p^\dag_{1}(p^\dag_{-1})$ are the creation operators of $J=1(-1)$ polaritons, $a_1$ and $a_2$ are the creation operators for the new eigenstates, and $\kappa=\frac{(V_{\textnormal{ex}})}{\sqrt{V_\textnormal{s}^2+V_\textnormal{ex}^2}}$ if $s_\textnormal{zl}=\frac{1}{2}$, or $\kappa=-\frac{(V_\textnormal{ex})}{\sqrt{V_\textnormal{s}^2+V_\textnormal{ex}^2}}$ if $s_\textnormal{zl}=-\frac{1}{2}$. The fraction of $J=1$ and $J=-1$ polariton in the lower energy state $a_1$ is $(1+\kappa)^2/2$ and $(1-\kappa)^2/2$ respectively. In the higher energy state $a_2$ the fraction of $J=1$ and $J=-1$ polaritons is $(1+\kappa)^2/2$ and $(1-\kappa)^2/2$ respectively. It follows that the fraction of $J=1(-1)$ polaritons is higher(lower) in the state $a_1$ than in $a_2$  when $s_\textnormal{ze}=-\frac{1}{2}$ $(\kappa<0)$. On the other hand if $s_\textnormal{ze}=\frac{1}{2}$, fraction of $J=1(-1)$ polaritons is lower(higher) in the state $a_1$ than in $a_2$ $(\kappa>0)$. These equations can be inverted to be written as:
\be 
p_1&=& \frac{1}{\sqrt{2}}\left (1-\kappa\right )a_1+\frac{i}{\sqrt{2}}\left(1+\kappa\right )a_{2},\nonumber\\
p_{-1}&=& \frac{1}{\sqrt{2}}\left (1+\kappa\right )a_1+\frac{-i}{\sqrt{2}}\left (1-\kappa\right )a_{2}.
\label{mix}
\ee
Consequently, a H-polarized measurement pulse pumps the lower energy state $a_1$ more than the higher energy state $a_2$, resulting in a asymmetric response from a two-sided cavity, shown in  Fig.3(a), for a typical $V_\textnormal{s}=0.15$ meV. \\
\indent
At $\delta=0$, the $a_1$ mode is excited faster than the $a_2$ mode, which in turn implies that the $\sigma^+$ light (corresponding to $J=1$ LPs) will be reflected more than the $\sigma^-$ light (corresponding to $J=-1$ LPs). Hence, at $\delta=0$, $I_\textnormal{D1}-I_\textnormal{D2}>0$, for both two-sided and single-sided cavity cases. Unlike the case when the H- and V-polarization modes were degenerate, the absorption response is non-zero in a single-sided cavity (Fig. 3(b)). Furthermore in a two sided cavity, at $\delta\sim V_\textnormal{s}$, the laser pulse is blue-detuned from the $a_1$ mode, but is on resonance with the $a_2$ mode. This means that although the pump rate for the $a_1$ mode is faster than for $a_2$, fewer a LPs in the $a_2$ mode are reflected out of the cavity. This makes the overall reflection of $a_1$ and $a_2$ modes to be the same, hence $|f_+|^2=|f_{-}|^2$ and $I_\textnormal{D1}-I_\textnormal{D2}=0$, at $\delta=V_\textnormal{s}$. Unlike the case when $V_\textnormal{s}=0$, $|f_+|$ and $|f_-|$ are non-zero for $\delta=0$ and hence the phase response does not vanish at zero detuning.
\begin{figure}
\includegraphics[width=4.5cm,height=4.5cm]{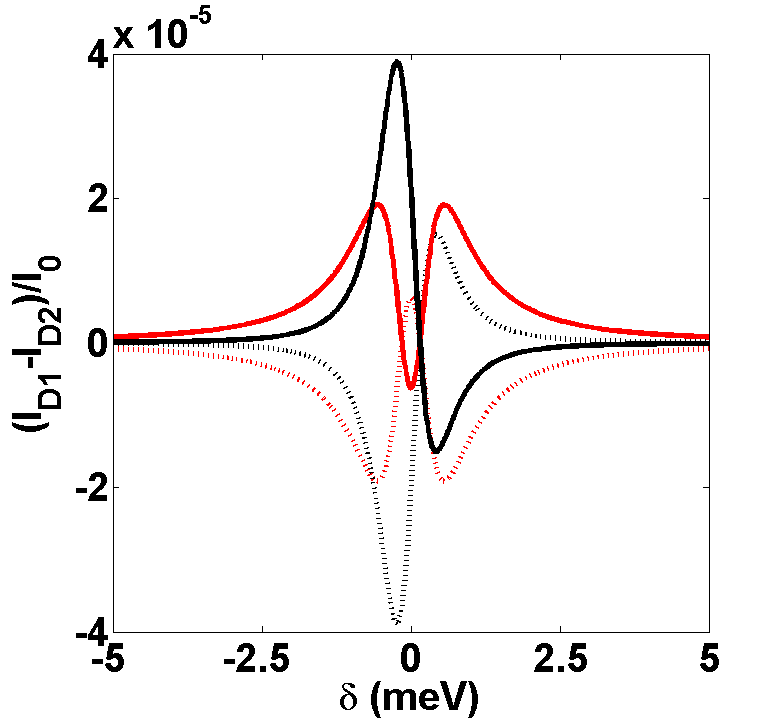}\includegraphics[width=4.5cm,height=4.5cm]{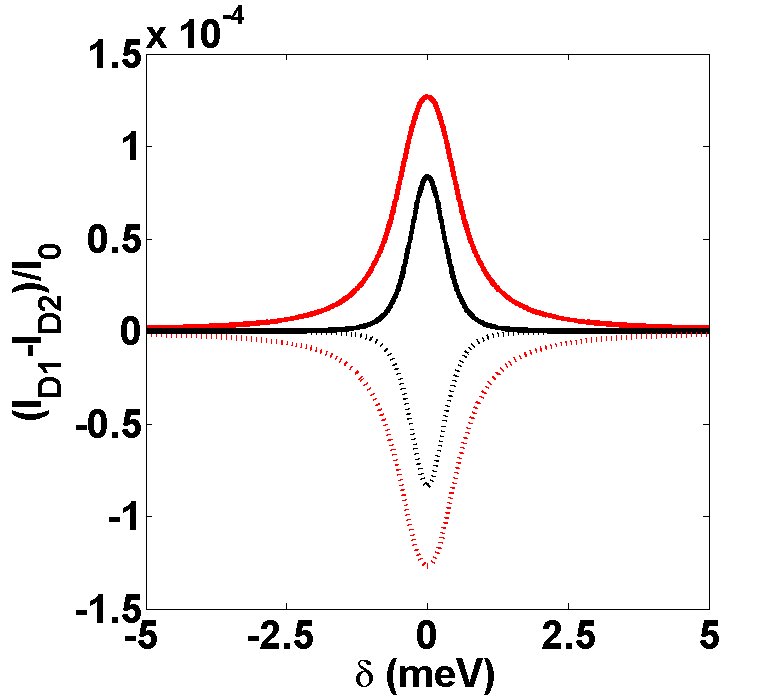}
\\
\hspace{1cm} (a)\hspace{4cm} (b)\\
\caption{Phase (red) and intensity (black) response when $V_s=0.15$ meV, (a) $\gamma_1=\gamma_2=0.5$ meV, (b) $\gamma_1=1$ meV, $\gamma_2=0$. The solid (dashed) lines represent the response with the QD electron spin is $s_{\textnormal{zl}}=\frac{1}{2}(-\frac{1}{2})$.}
\end{figure}

\section{QW Polariton Density}
If the polariton density ($n_\textnormal{pol}$)  becomes so high that the polariton-polariton interaction can no longer be ignored, then our analysis breaks down. As long as the interparticle separation is small compared to the scattering length approximated by the exciton Bohr radius, the polaritons behave as a weakly-interacting Bose gas \cite{hui}. At high densities, due to an increase in polariton repulsion, a blue shift of the LP branch is also observed \cite{resgeo}. We chose $n_\textnormal{pol}=5\times 10^9$ cm$^{-2}$, so that $n_\textnormal{pol} a^2_B\sim 0.005 \ll 1$. At this density the interparticle scattering can be ignored, as shown by the results in \cite{resgeo}.

\section{Error Estimates}

\subsection{Phonon assisted, spin-flip scattering rate}
\begin{figure}
\includegraphics[width=8cm,height=6cm]{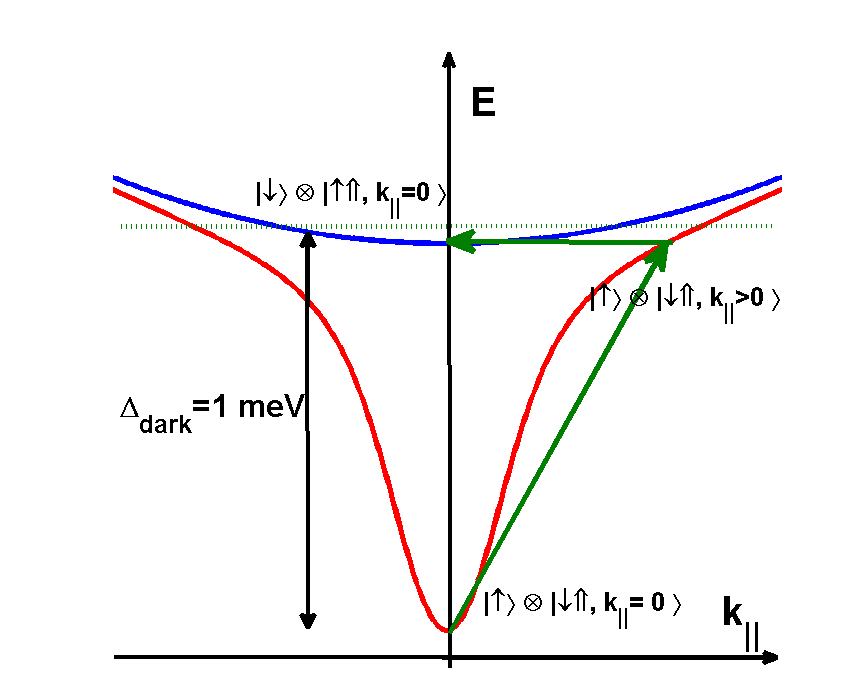}
\caption{Illustration (exaggerated) of the dark exciton (blue) and LP (red) dispersion curves. The $J=1$ LP at ${\bf{k}}_\|=\bf{0}$ ($\ket{\downarrow\Uparrow,k_\|=0}$) absorbs a phonon and is scattered to a state with momentum ${\bf{k}}'_\|$ ($\ket{\downarrow\Uparrow,{\bf{k}}'}_\|$), such that the energy of the LP at ${\bf{k}}'$ is equal to the dark exciton energy at ${\bf{k}}_\|={\bf{0}}$ ($\ket{\uparrow\Uparrow,k_\|=0}$). Now the QD electron spin can undergo a spin-flip via the Coulomb exchange interaction with the LP. The QD electron state with spin $s_\textnormal{ze}=\frac{1}{2}$ is written as $\ket{\uparrow}$ and with spin $s_\textnormal{ze}=-\frac{1}{2}$ is written as $\ket{\downarrow}$. For example the state $\ket{\uparrow}\otimes\ket{\downarrow\Uparrow,k_\|=0}$ means that the QD has spin $s_\textnormal{ze}=\frac{1}{2}$ and a LP has $J=1$ and ${\bf{k}}_\|=0$.}
\end{figure}
As was mentioned in the paper, the Coulomb exchange interaction comprises of spin conserving and spin-flip terms. The spin-flip term gives rise to read-out errors, which we estimate in this section. Figure 4 shows the exaggerated dispersion curves for the dark excitons and LPs. In our scheme, a long pump pulse coherently excites $N$ LPs at ${\bf{k}}_\|={\bf{0}}$. Suppose that at time $t=0$, the QD electron spin is $s_\textnormal{ze}=\frac{1}{2}$ ($\ket{\uparrow}$). The spin can flip via the spin-flip term in the Coulomb exchange interaction Hamiltonian, scattering the $J=1$ LP at ${\bf{k}}_\|={\bf{0}}$ ($\ket{\downarrow\Uparrow,{\bf{k}}_\|={\bf{0}}}$) to a $J=2$ dark exciton ($\ket{\uparrow\Uparrow,{\bf{k}}_\|={\bf{0}}}$). As shown in Fig. 4, the dark exciton is blue detuned from the LP. This direct scattering does not conserve energy and hence is prohibited. However, at finite temperature $T$, the ${\bf{k}}_\|={\bf{0}}$ LP can get scattered by a phonon into a higher energy state with in-plane momentum ${\bf{k}}'_\|$ ($\ket{\downarrow\Uparrow,{\bf{k}}'_\|}$). If the polariton in this mode has energy $E_\textnormal{LP}({\bf{k}}'_\|)$ equal to that of the dark exciton at ${\bf{k}}''_\|$, then the spin-flip exchange interaction becomes possible. For example, the smallest in-plane momentum of the LP, for which an energy conserving spin-flip process takes place is ${\bf{k}}'_{\|,0}$, such that $E_\textnormal{LP}({\bf{k}}'_{\|,0})=\Delta_\textnormal{dark}$. The resulting read-out error depends on two rates: (i) the rate at which the LPs at ${\bf{k}}_\|=0$ are scattered into modes with $|{\bf{k}}'_\| |\geq |{\bf{k}}'_{\|,0}|$ by phonon absorption and (ii) the rate at which the QD electron spin $s_\textnormal{ze}=\pm \frac{1}{2}$ and the $J=\pm 1$ LP at ${\bf{k}}'_\|$ scatter via the spin-flip exchange interaction into $s_\textnormal{ze}=\mp \frac{1}{2}$ and the $J=\pm 2$ dark exciton at ${\bf{k}}''$ such that  $E_\textnormal{LP}({\bf{k}}'_\|)=E_\textnormal{dark}({\bf{k}}''_\|)$. Because of high density of states of the dark-exciton, the second process is very fast ($\sim$ 10 ns), so the fidelity of our measurement scheme is governed by the slower phonon absorption process, which we estimate next. 
\\
\indent
 The rate at which a LP at ${\bf{k}}_\|={\bf{0}}$ absorbs a phonon with in-plane momentum ${\bf{q}}$, and $z$ axis momentum $q_z$  $({\bf{q}},q_\textnormal{z})$ and is scattered to a LP at momentum ${\bf{k}}'_\|$ is:
\be
W_{{\bf{0}}\rightarrow {\bf{k}}'_\|}&=&\frac{2\pi}{\hbar}\sum_{q_\textnormal{z}}|r_0|^2|r_{{\bf{k}}'_\|}|^2|\bra{{\bf{k}}'_\|,n_{{\bf{q}},q_\textnormal{z}}}\textnormal{H}_{\textnormal{exc-ph}}\ket{{\bf{0}},n_{{\bf{q}},q_\textnormal{z}}+1}|^2\nonumber\\
&\times& \delta(E_\textnormal{LP}({\bf{k}}'_\|)-E_\textnormal{ph}({\bf{q}},q_\textnormal{z})-E_\textnormal{LP}({\bf{0}})-\delta),
\label{kl}
\ee
where, $\textnormal{H}_{\textnormal{exc-ph}}$ is the exciton-phonon interaction Hamiltonian \cite{pie}, $r_0,r_{{\bf{k}}'}$ are the exciton Hopfield coefficients, $E_\textnormal{LP}({\bf{k}}')$, $E_\textnormal{ph}({\bf{q}},q_\textnormal{z})$ are the energies of the LP, phonon, $\delta$ is the detuning the probe pulse from the LP resonance at ${\bf{k}}={\bf{0}}$ and $n_{{\bf{q}},q_\textnormal{z}}$ is the Bose distribution function of phonons at temperature $T$. It is important to note that since the LPs are virtually excited, their energy at ${\bf{k}=0}$ is the same as the energy of the probe laser and hence the extra term of $\delta$ is the argument of the Dirac-Delta function in the expression above. For example, in order to get maximum signal when using a symmetric two-sided cavity the probe laser (at frequency $\omega_\textnormal{P}$) must be detuned by $\delta\sim 0.3$ meV from the LP resonance (Fig. 3(a) in the main paper). Hence, the LPs excited by the probe pulse have energy = $\hbar\omega_\textnormal{P}$. If the longitudinal velocity of acoustic phonons in the QW is $u$, then, $E_\textnormal{ph}({\bf{q}},q_\textnormal{z})=\hbar u\sqrt{{\bf{q}}^2+q^2_\textnormal{z}}$. If the Rabi splitting of the polaritons in the QW is $g$, then:
\be 
E_\textnormal{LP}({\bf{k}}'_\|)&=&\frac{E_\textnormal{cav}({\bf{k}}'_\|)+E_\textnormal{exc}({\bf{k}}'_\|)}{2}\nonumber\\
&-&\frac{\sqrt{g^2+(E_\textnormal{cav}({\bf{k}}'_\|)-E_\textnormal{exc}({\bf{k}}'_\|))^2}}{2}.
\ee
Here, $E_\textnormal{cav}({\bf{k}}'_\|)$ and $E_\textnormal{exc}({\bf{k}}'_\|)$ are the cavity photon and exciton energies respectively.
The exciton-phonon Hamiltonian is given by:
\be 
\textnormal{H}_{\textnormal{exc-ph}}=\sum_{q_\textnormal{z}}\sum_{{\bf{q}},{\bf{k}}'_\|}G({\bf{q}},q_\textnormal{z})\delta_{{\bf{k}}'_\|,{\bf{q}}}(c_{{\bf{q}},q_\textnormal{z}}-c^\dag_{-{\bf{q}},q_\textnormal{z}})b^\dag_{{\bf{k}}'_\|}b_{0},\nonumber\\
\ee
where, $c$, $b$ are the annihilation operators of the phonons, excitons and $G({\bf{q}},q_\textnormal{z})$ contains all the interaction terms. Note that because of translational invariance only along the plane of the QW, conservation of momentum is valid only for in-plane momentum.
\be 
G({\bf{q}},q_\textnormal{z})&=&i\sqrt{\frac{\hbar \sqrt{{\bf{q}}^2+q^2_\textnormal{z}}}{2\rho V u}}\nonumber\\
&\times&[a_\textnormal{e}I^\|_\textnormal{e}(|{\bf{q}}|)I^\perp_\textnormal{e}(q_\textnormal{z})-a_\textnormal{h}I^\|_\textnormal{h}(|{\bf{q}}|)I^\perp_\textnormal{h}(q_\textnormal{z})],
\label{kl2}
\ee
where, $\rho$ is the mass density, $V$ is the quantization volume, $a_\textnormal{e(h)}$ are the electron, hole deformation potential experimentally measured and $I^{\|(\perp)}_\textnormal{e(h)}$ are the superposition integrals of the electron, hole part of the exciton and phonon wavefunctions in the in-plane and $z$ direction:
\be 
I^\|_\textnormal{e(h)}(|{\bf{q}}|)&=&\left[1+\left(\frac{m_\textnormal{h(e)}}{2M}|{\bf{q}}|a_\textnormal{B}\right)^2\right]^{-3/2}, \nonumber\\
I^\perp_\textnormal{e}(q_\textnormal{z})&=&\int |g(z)|^2e^{iq_\textnormal{z}z}dz, \textnormal{ and } I^\perp_\textnormal{h}(q_\textnormal{z})=\int |h(z)|^2e^{iq_\textnormal{z}z}dz.\nonumber\\
\label{kl3}
\ee
 The error rate depends on the scattering of LPs into states with $|{\bf{k}}'_\||\geq |{\bf{k}}'_{\|,0}|$:
\be 
\Gamma=\sum_{|{\bf{k}}'|\geq |{\bf{k}}_0|}W_{{\bf{0}}\rightarrow {\bf{k}}'_\|}
\ee
 Using Eqns. \ref{kl}-\ref{kl3} with $a_\textnormal{e}=-7$ eV, $a_\textnormal{h}=2.7$ eV,  $T=1.5$ K and $g=2$ meV, we get $\Gamma=30$ s$^{-1}$ when using a single sided cavity so that $\delta=0$ to obtain maximum signal strength. In a two-sided cavity case, $\delta=0.3$ meV and $\Gamma= 0.2$ s$^{-1}$. \\
\indent
Note that, it is possible for UPs to get scattered to LPs by phonon assisted spontaneous emission or absorption. However, since the probe pulse is red detuned from the UP resonance only absorptive scattering to LPs is possible. In the case of a single-sided cavity, $\delta=0$ and $N=2000$ LPs and $N=110$  UPs are excited. Hence the total phonon-assisted scattering rate = $\Gamma^\textnormal{dark}=63300$ s$^{-1}$. In  two-sided cavity, $\delta=0.3$ meV and $N=2000$ LPs and $N=90$  UPs are excited. Hence the total phonon-assisted scattering rate = $\Gamma^\textnormal{dark}=418$ s$^{-1}$.

\subsection{Localized electron qubit-QW hole radiative recombination rate}
As long as the probe beam is pumping exciton-polaritons in the QW, there is finite probability for the localized electron spin qubit to radiatively recombine with one of the holes. For a single exciton in the QW, the initial state consists of an electron with the QD overlap wavefunction in the conduction band, and another electron in the conduction band, and a hole in the valence band with the overlap wavefunction of the QW exciton. After the QD electron recombines with the QW hole that originally had formed the exciton, an electron will be left behind, and we assume it will take on the QD wavefunction. The dipole transmission matrix element for this transition from initial state $\ket{i}$ to final state $\ket{f}$ is:
\be 
&&\langle f|u(\vec{k})\vec{p}\cdot \hat{\epsilon}_{\vec{k}}| i \rangle \nonumber\\
&=& \frac{\vec{p}_\textnormal{cv}\cdot \hat{\epsilon}_{\vec{k}}}{\sqrt{DL'}}\int \int \phi^*(\vec{r}_\textnormal{e})e^{i\vec{k}\cdot \vec{r}_\textnormal{d}}\phi(\vec{r}_\textnormal{d})\psi(\vec{r}_\textnormal{e},\vec{r}_\textnormal{d})  d\vec{r}_\textnormal{e} d\vec{r}_\textnormal{d},\nonumber
\ee
where $D$, is the quantization area of the field along the plane of the QW, $L'$ is the quantization length of the field perpendicular to the plane of the QW, $\vec{p}_\textnormal{cv}$ is the electric dipole moment for the transition of an electron from the valence to the conduction band, and $\hat{\epsilon}_{\vec{k}}$ is the unit vector in the direction of the field with wavevector $\vec{k}$.
Using the wavefunctions listed in sections I and II, and employing the dipole approximation, one can simplify the above equation to:
\be 
\langle f|u(\vec{k})\vec{p}\cdot \hat{\epsilon}_{\vec{k}}| i \rangle =\frac{\vec{p}_\textnormal{cv}\cdot \hat{\epsilon}_{\vec{k}}}{\sqrt{DL'}} \Gamma \phi(0)e^{-|{\bf{k}}_\||^2a^2/4}, 
\ee
where
\be
\Gamma = N_\textnormal{m}^2 M^2I_0^2\frac{1}{\sqrt{A}}I_0^2\pi^2 \frac{a^2}{1/a^2+8/a_\textnormal{B}^2},\quad \phi(0)=\sqrt{\frac{2}{\pi}}\frac{2}{a_\textnormal{B}},\nonumber
\ee
and $\vec{k}=({\bf{k}}_\|,k_\textnormal{z})$. Hence the oscillator strength is \cite{sug}:
\be 
f_{\vec{k}}&=&\frac{|\vec{p}_\textnormal{cv}\cdot \hat{\epsilon}_{\vec{k}}|^2}{Dm_0\omega_\textnormal{x}} \Gamma^2 |\phi(0)|^2e^{-|{\bf{k}}_\||^2a^2/2}\nonumber\\
&=&\left(\left(1+\frac{k_{\textnormal{z}}^2}{k^2}\right )f_\|+\left(\frac{k_\textnormal{x}^2}{k^2}\right )f_\bot\right)\frac{\Gamma^2}{D}e^{-k_\textnormal{x}^2a^2/2}
\ee
where
\be
f_\|=\frac{M^2(e_\|)p_\textnormal{cv}^2|\phi(0)|^2}{m_0\omega_x},\quad f_\bot =\frac{M^2(e_\bot)p_\textnormal{cv}^2|\phi(0)|^2}{m_0\omega_\textnormal{x}}\nonumber
\ee
Here $\omega_x$ is the exciton recombination energy, $M(e_\|)$ and $M(e_\bot)$ are the polarization factors parallel and perpendicular to the QW. Thus the decay rate $\tau_D^{-1}$ is:
\be 
\tau_D^{-1}&=&\frac{\pi e^2\omega_\textnormal{x}}{m_0\epsilon_0 c n L'}\sum_{\vec{k}} \frac{f_{\vec{k}}}{\omega_{\vec{k}}}\delta(k_\textnormal{x}-k)\nonumber\\
&\approx&\frac{2e^2n\omega_\textnormal{x}^2}{m_0\epsilon_0 c^3}\frac{\Gamma^2}{2\pi}\bar{f}, \quad \bar{f}=\frac{2}{3}f_\|+\frac{1}{3}f_\bot=\frac{p_\textnormal{cv}^2|\phi(0)|^2}{m_0\omega_\textnormal{x}}\nonumber\\
&=&\tau_0^{-1}\frac{n^2\omega_\textnormal{x}^2}{c^2}\frac{\Gamma^2}{2\pi}\frac{2}{3}\nonumber
\ee
where $\tau_0=23$ ps is the decay rate for free excitons with zero in-plane momentum. Thus $\tau_D\sim 0.39$ s, and if there are $N$ excitons in the QW, $\tau_D\sim 0.39/N$ s.

\subsection{Shot Noise Error Estimation}
In a photon counting experiment, if the signal is a coherent state $\ket{\alpha}$, then the inherent quantum noise due to the Poissonian photon number statistics of the coherent state is $\Delta(n)$=$|\alpha|$. This is called the shot noise. In our QND measurement setup, we have coherent states with photon fluxes $I_\textnormal{D1}$ and $I_\textnormal{D2}$ incident on detectors D1 and D2 respectively. After a measurement time $\tau_\textnormal{meas}$, the total expected number of photons incident on detector D1 is $n_1=\tau_\textnormal{meas}I_\textnormal{D1}$ and on D2 is $n_2=\tau_\textnormal{meas}I_\textnormal{D2}$. Our measurement signal is $N_1-N_2$, where $N_1$ and $N_2$ are the photon counts actually measured by detectors D1 and D2. For example, in a single-sided cavity at $\delta=0$, the photon flux of the measurement pulse, so that 2000 LPs are generated in the QW, is $I_0=5\times 10^3$ ps$^{-1}$. If $s_\textnormal{ze}=\frac{1}{2}$, the photon fluxes incident on D1 and D2 are $I_\textnormal{D1}=225.414$ ps$^{-1}$ and $I_\textnormal{D2}=224.586$ ps$^{-1}$ respectively. Hence, for a measurement time of $\tau_\textnormal{meas}=8$ ns (table 1 in the paper), the average photon counts measured by D1 and D2 are $n_1=1803312$ and $n_2=1796688$.  By a similar analysis we estimate that, if $s_\textnormal{ze}=-\frac{1}{2}$, then on average D1 will record $1796688$ counts and D2 will record $1803312$ counts. Thus the difference in the detector counts $n_1-n_2$ is a measure of the spin state of the QD electron. However, since the input to the detectors are coherent states, the difference in their counts will follow Poissonian statistics, with mean photon number = $\pm (n_1-n_2)$ and standard deviation $\sigma=\sqrt{n_1^2+n_2^2}$ (Fig. 4). Hence, the probability to determine the spin state of the QD electron as $s_\textnormal{ze}=-\frac{1}{2}$, when it was actually $s_\textnormal{ze}=\frac{1}{2}$ is equal to the probability that the difference in the counts of the two detectors was negative. This error probability is equal to the area of the blue shaded region in Fig. 4. Similarly, the probability of determining $s_\textnormal{ze}=\frac{1}{2}$, when actually $s_\textnormal{ze}=-\frac{1}{2}$, is equal to the area of the red shaded region in Fig. 4.
Since the mean photon numbers are large, we approximate the Poisson distribution as a Gaussian. The probability of error is then computed as:
\be 
P_\textnormal{e}^{\textnormal{sn}}&=&\frac{1}{\sqrt{2\pi}\sigma}\int_{-\infty}^0 e^{-\frac{(x-(n_1-n_2))^2}{2\sigma}} dx\nonumber\\
&+&\frac{1}{\sqrt{2\pi}\sigma}\int_{0}^\infty e^{-\frac{(x-(n_2-n_1))^2}{2\sigma}} dx\nonumber\\
&=&\textnormal{erfc}\left (\frac{|n_1-n_2|}{\sqrt{2}\sigma}\right)
\ee
For the example of a single-sided cavity introduced previously, $n_1-n_2=6624$ and $\sigma=1897.4$, hence $P_\textnormal{e}^{\textnormal{sn}}\sim 4\times 10^{-4}$.

\begin{figure}
\includegraphics[width=4.5cm,height=4cm]{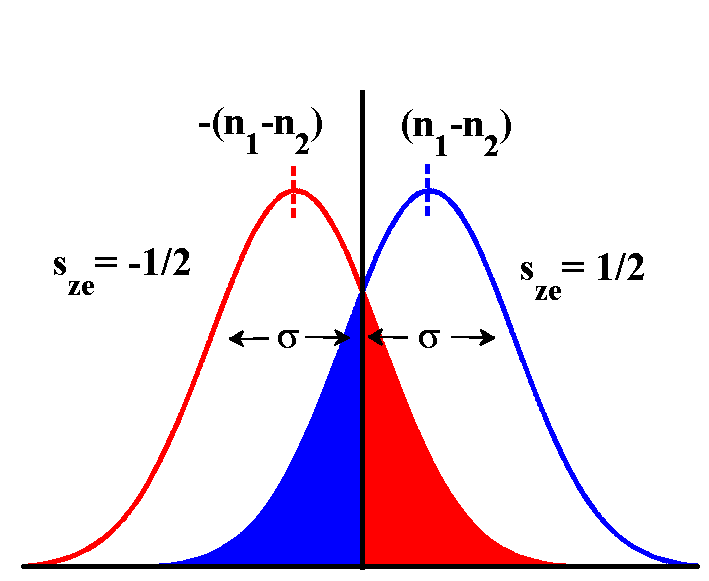}
\caption{Statistics of the difference in the photon counts of two detectors. If $N_1-N_2>0(<0)$, then $s_\textnormal{ze}=\frac{1}{2}(-\frac{1}{2})$. The probability that an error occurs is the area of the shaded region.}
\end{figure}

\section{Cavity Parameters}
If the reflectivites of the top and bottom DBR mirrors are $r_1$ and $r_2$ respectively, then the rate of decay from the top and bottom mirrors are \cite{book}:
\be
\gamma_1\sim\frac{L_\textnormal{c}n_\textnormal{c}}{c(1-r_1)}, \quad \gamma_2\sim\frac{L_\textnormal{c}n_\textnormal{c}}{c(1-r_2)},
\ee
where $L_\textnormal{c}$ is the effective cavity length, $n_\textnormal{c}$ is the refractive index of the cavity and $c$ is the vacuum speed of light. The above expression is valid when $r_1,r_2\sim 1$. In such a cavity, the unique spot size ($\pi R^2$) in which the polaritons are excited is:
\be 
R=\sqrt{\frac{\lambda L_\textnormal{c}}{\pi(1-r_1r_2)}},
\ee
where $\lambda$ is the wavelength of the cavity photon. Hence, in a two-sided symmetric cavity if $r_1=r_2=99.9\%$, then $\gamma_1=\gamma_2=0.5$ meV and $R=3.6$ $\mu$m.

\end{document}